\documentstyle[twocolumn,prl,aps]{revtex}

\begin{document}
\title{Deep Inelastic Neutron Scattering determination of the single particle
kinetic energy in solid and liquid $^3$He}
\author{R. Senesi}
\address{Istituto Nazionale per la Fisica della Materia,\\
UdR Roma Tor Vergata, via della Ricerca Scientifica 1, 00133 Roma, Italy.}
\author{C. Andreani}
\address{Dipartimento di Fisica, Universit\`{a} degli Studi di Roma \\
Tor Vergata and\\
Istituto Nazionale per la Fisica della Materia, via della Ricerca\\
Scientifica 1, 00133 Roma, Italy.}
\author{D. Colognesi}
\address{Consiglio Nazionale delle Ricerche, Gruppo Nazionale 
Struttura della Materia
and Istituto Nazionale per la Fisica della Materia, UdR Tor Vergata, Roma,
Italy.}
\author{A. Cunsolo}
\address{Istituto Nazionale per la Fisica della Materia, \\
UdR Roma Tre, via della Vasca Navale 84,\\
00146, Roma, Italy.}
\author{M. Nardone}
\address{Dipartimento di Fisica Universit\`{a} degli Studi di Roma Tre and \\
Istituto Nazionale per la Fisica della Materia,\\
via della Vasca Navale 84, 00146, Roma, Italy.}
\date{\today}
\maketitle

\begin{abstract}
\noindent For the first time, an experimental determination of the
single-particle mean kinetic energies for $^{3}$He along the $T=2.00$ K
isotherm in the dense liquid and in the solid hcp and bcc phases is
reported. Deep Inelastic Neutron Scattering measurements at exchanged
wavevectors ranging from $90.0$ \AA $^{-1}$ to $140.0$ \AA$^{-1}$ have been
performed in order to evaluate, within the framework of the Impulse
Approximation, the molar volume dependence of the kinetic energy. The
results are found in excellent agreement with recent Diffusion Monte Carlo
simulations.\newline
\end{abstract}

\indent
The study of the solid helium isotopes is of basic interest in condensed
matter physics since these systems represent the prototype of a quantum
solid. For this reason considerable efforts have been addressed to the
understanding of their microscopic static and dynamical properties from both
the experimental and the theoretical points of view \cite{Glyde,Dobbs}. As
far as the dynamics is concerned neutron spectroscopy can provide unique
information about both collective and single-particle excitations in these
systems. In particular, in the case of solid $^{4}$He, {\it Inelastic
Neutron Scattering} (INS) from collective modes has been extensively
employed to test the inclusion of anharmonic contributions in theories, such
as, for instance the {\it self-consistent phonon} approach \cite{Glyde}.
More recently {\it Deep Inelastic Neutron Scattering} (DINS) has provided
information on the single-particle momentum distribution function,
particularly on the mean kinetic energies, which have been successfully
compared with the most advanced ground-state simulation techniques \cite
{Momentum,Simmons,Zoppi,Ceperley,Moleko,Zoppi1}. On the contrary $^{3}$He
has been by far less investigated mainly because of its large neutron
absorption cross section. The only exceptions are some solid phase neutron
diffraction measurements \cite{Benoit,Bossy,Siemensmeyer} and, more
recently, some DINS measurement performed in the liquid phase which have
proven to be a very sensitive test of the single-particle properties derived
from theoretical models \cite{Sokol,Mook,Azuah,Moroni,Casulleras,ceperley92}.%
\newline
\newline
\indent

In this letter we report the first experimental determination of
single-particle mean kinetic energies $\left\langle E_{K}\right\rangle $ in
the solid, bcc and hcp phases, and in the high density liquid $^{3}$He, by
means of DINS. This technique is at present the only one which allows direct
access to single-particle dynamical properties, such as momentum
distribution, $n(\vec{p})$, and in particular $\left\langle
E_{K}\right\rangle ,$ \cite{evs} exploiting the large values of momentum
exchange and energy transfer in the scattering process \cite{Momentum}. As a
matter of fact, under these circumstances, the {\it Impulse Approximation}
(IA) yields a scattered intensity, determined by the single particle
dynamical structure factor $S(\vec{q},\omega ),$ which turns out to be given
by:
\begin{equation}
S_{IA}(\vec{q},\omega )=\int n(\vec{p})~\delta \left( \omega -\frac{\hbar
q^{2}}{2M}-\frac{\vec{q}\cdot \vec{p}}{M}\right) d\vec{p},
\end{equation}
where $M$ is the sample atomic mass. This scattering law can be then
expressed in terms of a scaling function: $F(y,\hat{q})=\frac{\hbar q}{M}$ $%
S_{IA}(\vec{q},\omega )$, where $y=\frac{M}{\hbar q}\left( \omega -\frac{%
\hbar q^{2}}{2M}\right) $ is the West scaling variable \cite{Watson}. The
function $F(y,\hat{q})$, often referred to as the {\it Neutron Compton
Profile} (NCP) \cite{Watson}, represents the probability density
distribution of $y$, the atomic momentum component along the direction of
momentum transfer $\hat{q}$.\newline
\newline
\indent
The present DINS experiment has been performed on the eVS spectrometer, an
inverse-geometry instrument operating at the ISIS pulsed neutron source
(Rutherford Appleton Laboratory, Chilton, Didcot, UK) \cite{evs}, where an
intense flux of incident neutron in the $1-100$ eV spectral range is
available. It has to be stressed that, owing to the already quoted
absorption problems, measurements on solid $^{3}$He require a careful choice
of the experimental parameters and of the scattering geometry. Indeed the
neutron absorption cross section for $^{3}$He atoms has been estimated to be
$\sigma _{a}=$ $5333$ barn for $25$ meV neutrons \cite{mughabghab}, about
three orders of magnitude larger than the total scattering cross section $%
\sigma _{sc}=(6.85$ $\pm $ $0.12)$ barn \cite{Guckelsberger}, and to follow
the usual $1/v$ law, $v$ being the neutron velocity. In order to minimize
the absorption of the sample it is therefore important to perform the
experiment using the highest available incident neutron energies. In
addition the use of a backscattering geometry is mandatory in order to
achieve the high momentum transfer required by the IA. The inelastic neutron
spectrum has been determined using the filter difference technique measuring
the time of flight of the neutrons absorbed by the $4.908$ eV resonance of
the Au foil filter \cite{evs}. The scattered neutrons have been detected
using scintillator detectors placed in the angular range $100^{o}<$ $2\theta
<148^{o}$, thus yielding a momentum transfer ranging from $90.0$ \AA $^{-1}$
to $140.0$ \AA $^{-1}$ and an energy transfer from $5$ eV to $13$ eV \cite
{evs}. This yields an average ratio between the absorption and scattering
cross section of about $30$.\newline
\newline
\indent
The experiment has been performed at a constant temperature of $2.00$ K
varying the applied pressure in order to obtain a high density liquid
sample, a body centered cubic (bcc) sample and an hexagonal close packed
(hcp) sample of $^{3}$He (see Tab. I) . The samples were contained in a
cylindrical annular aluminum-alloy can (inner diameter of the sample $\simeq
$ 20 mm, outer diameter $\simeq $ 22 mm, wall thickness $\simeq $ 1 mm)
inserted in a liquid helium flow cryostat. The $^{3}$He gas was condensed in
the cell kept, through all the experiment, at a constant temperature of 2.00
K $\pm $\ 0.01 K as measured by two Ge resistance thermometers located in
the upper and lower ends of the sample cell. Pressure was subsequently
increased in order to explore the isotherm using an inert gas pressure
intensifier. The molar volumes of the samples in the present experiment have
been determined as follows: for the liquid sample, the value from Ref. \cite
{gibbons} corresponding to the measured pressure of $p=$50 bar has been
used, while for the bcc and hcp solid samples the molar volume has been
experimentally derived from the measured transmission of the incident
neutron beam, relatively to that of the liquid sample, performed at three
different energy values. Transmissions were obtained from two proportional
monitor detectors placed before and after the sample, respectively. The time
of flight calibration of the two monitors has been performed exploiting
three resonances (namely at 6.67 eV, 36.68 eV, 66.02 eV) of a $^{238}$\ U
foil placed before both monitors with no sample in the beam. In the present
DINS experiment this molar volume determination procedure appears more
reliable than the usual diffraction pattern analysis \cite{Zoppi1}, due to
the high absorption of the samples and the high momentum transfer values.
Experimental values of the molar volume in the three thermodynamic states
are reported in Table I. The time-of-flight spectra were corrected for
detector efficiency and normalized to the monitor counts by using standard
routines available on eVS \cite{evs}. Thanks to the high absorption cross
section of $^{3}$He samples, the estimated multiple scattering contribution
turned out to be negligible \cite{Sears}. As far as the absorption
correction is concerned, it has been evaluated through a Monte Carlo
simulation \cite{Sokol,3he4he,3he4he2001}. The fixed-angle experimental
resolution, $R_{n}(y),$ determined for each $n^{th}$ detector through an
experimental standard calibration using a lead sample , was found, as in
previous {\it eVS } measurements on $^{4}$He and $^{3}$He \cite
{3he4he,3he4he2001,Albergamo,superhe}, to be well described by a Voigt
function. This calibration ensures a reliable estimate of the line-shape
resolution function \cite{FieldingNIM}. Experimental spectra from different
detectors have been transformed into $y$ space and described in terms of the
response scaling function, $F(y,q)$, where the dependence on the direction
of momentum transfer $\hat{q}$ has been omitted due to the liquid and
polycrystalline nature of the samples with no preferred orientation. The $q$%
-dependence of $F(y,q)$\ arising from deviations to the IA is generally
described in terms of the {\it Final State Effects} (FSE) \cite{Glyde,Sears1}%
. Strictly speaking this dependence vanishes in principle only in the
asymptotic limit $q\rightarrow \infty $\ (performed keeping $y=$\ const).
\cite{Glyde,Sears1}. However it has been shown, both theoretically \cite
{Rinat} and experimentally \cite{superhe,Bill}, that for liquid $^{4}$He in
the high $q$ range accessed by eVS, FSE are negligible and do not affect
significantly the peak shape. Furthermore a self-consistent phonon
evaluation of the lattice density of states \cite{Glyde1} confirms that, for
solid $^{4}$He and $^{3}$He and for $q$\ larger than 30 \AA $^{-1}$, these
effects are almost negligible and shows that their leading term is odd in $y$%
. Therefore we have performed a $y$-symmetrization \cite{Sears1} on present
data in order to cope with the presence of small FSE before performing the
analysis within the IA framework\cite{superhe}. Previous DINS measurements
on solid $^{4}$He confirm the reliability of such a procedure \cite
{Simmons,Zoppi,Ceperley,Zoppi1}. Following this procedure a single function,
$\overline{F}(y)$, averaged over all the 32 detectors is derived. It is
reported, for bcc and hcp phases, in Fig. 1 where the experimental
resolution function, $\overline{R}(y)$, obtained averaging the single
detector experimental resolution $R_{n}(y),$\ is also reported (dotted
line). The values for $\left\langle E_{K}\right\rangle $\ are obtained
assuming a Gaussian function for the $n(p)$\ and exploiting the second
moment sum rule for $\overline{F}(y)$ \cite{Glyde,Watson}:
\begin{equation}
\int_{-\infty }^{\infty }y^{2}\overline{F}(y)~dy=\sigma _{y}^{2}=\frac{2M}{%
3\hbar ^{2}}\left\langle E_{K}\right\rangle
\end{equation}
where $\sigma _{y}$ is the standard deviation of $\overline{F}(y)$.\newline
\newline
\indent
For solid samples, the choice of a Gaussian function for the $n(p)$ , and
hence for $\overline{F}(y),$ can be justified through the use of the central
limit theorem \cite{Sears2}, even in presence of strong anharmonicities, and
it has already been successfully used in solid $^{4}$He \cite
{Simmons,Zoppi,Ceperley,Zoppi1}. For the liquid sample we stress that, even
at low densities $^{3}$He where deviations from Gaussian shape are expected
to be relevant, previous attempts to extract $\left\langle
E_{K}\right\rangle $, through both Gaussian and non-Gaussian $n(p)$, have
provided the same results within the experimental uncertainties \cite{Azuah}%
. In the present work, values of $\sigma _{y}$ are obtained fitting the
experimental data with a Voigt function, resulting from the convolution of
this Gaussian representing $\overline{F}(y)$ and the experimental resolution
function, $\overline{R}(y)$. An example of the fit on the experimental data
is shown in Fig. 1 (full line) while the values of the mean kinetic energy
derived are reported in Tab. I. In Fig. 2 the molar volume dependence of $%
\left\langle E_{K}\right\rangle $ is plotted together with the same quantity
evaluated by both Diffusion Monte Carlo (DMC) calculations \cite{Moroni00}
and by the self-consistent phonon method \cite{Moleko88}. From Fig. 2 we
observe that the experimental values of $\left\langle E_{K}\right\rangle $
are in a remarkable agreement with the DMC calculations for both liquid and
solid phases, while they lie systematically above the results of the self
consistent phonon approach. It can be noted that since in the latter
approach the anharmonicity is accounted for only through the introduction of
a cubic term, it is suggested that higher order anharmonic terms need to be
considered.

\indent
In conclusion for the first time experimental determinations of the
single-atom mean kinetic energy in solid and high density liquid $^{3}$He
are reported. The experimental method employed, namely Deep Inelastic
Neutron Scattering of epithermal neutrons, allowed to overcome the
difficulties arising both from the strong neutron $^{3}$He absorption and
from the presence of FSE, which render the extraction of $\left\langle
E_{K}\right\rangle $ particularly difficult \cite{Bill}. The present DINS
results, carefully analyzed taking into account an accurate experimental
determination of the spectrometer resolution, are in excellent agreement
with the most recent Diffusion Monte Carlo simulations of $\left\langle
E_{K}\right\rangle $ \cite{Moroni00}. Thus the experimental results provide
a valuble test on the quantitative realibilty of the interparticle
interactions adopted in the simulation. In our opinion it would be
worthwhile to extend experimental determinations of the $\left\langle
E_{K}\right\rangle $ to other thermodynamic states of solid $^{3}$He and
helium isotopic mixtures, in order to obtain a global picture of the density
and temperature dependence of kinetic energy in these systems.

{\it The authors gratefully acknowledge Prof. W. G. Stirling and Dr. S.
Moroni for the stimulating and interesting discussions and Mr. A.
Pietropaolo and the ISIS User Support Group for the valuable technical
support during experimental measurements.}

\section*{Tables}

TABLE I: \newline
\newline
\begin{tabular}{|c|c|c|}
\hline
Phase & $v$ (cm$^{3}/$mole) & $\left\langle E_{K}\right\rangle $ (K) \\
\hline\hline
L & 23.81$\pm $0.01 & 23.8$\pm $2.8 \\ \hline
bcc & 20.10$\pm $0.10 & 33.2$\pm $3.0 \\ \hline
hcp & 18.75$\pm $0.30 & 36.5$\pm $4.7 \\ \hline
\end{tabular}

\section*{Table Captions}

\noindent TAB. I: Molar volume and experimental single particle mean kinetic
energy, $\langle E_{K}\rangle $, in solid and liquid $^{3}$He, with their
estimated standard deviations. The two values of the molar volume in the
solid phase have been determined experimentally (see main text), while value
for the liquid is from Ref. \cite{gibbons}

\section{Figure Captions}

\noindent FIG. 1: Experimental response functions $\overline{F}(y)$ for : a)
the hcp $^{3}$He sample at the molar volume $v$ = 18.75 cm$^{3}/$mole and b)
the bcc sample at the molar volume $v$ = 20.10 cm$^{3}/$ mole. The $%
\overline{F}(y)$ was obtained symmetrizing the individual detector response
function before averaging over the 32 detectors (see main text). Best fit $%
\overline{F}(y)$\ (see main text) and the experimental resolution function, $%
\overline{R}(y)$\ are also reported as full and dotted lines respectively.%
\newline
\newline
FIG. 2: $^{3}$He mean kinetic energy, $\langle E_{K}\rangle $, as a function
of the molar volume: experimental values (solid circles), DMC values \cite
{Moroni00} (open circles) and self-consistent phonon method \cite{Moleko88}
(dashed line).

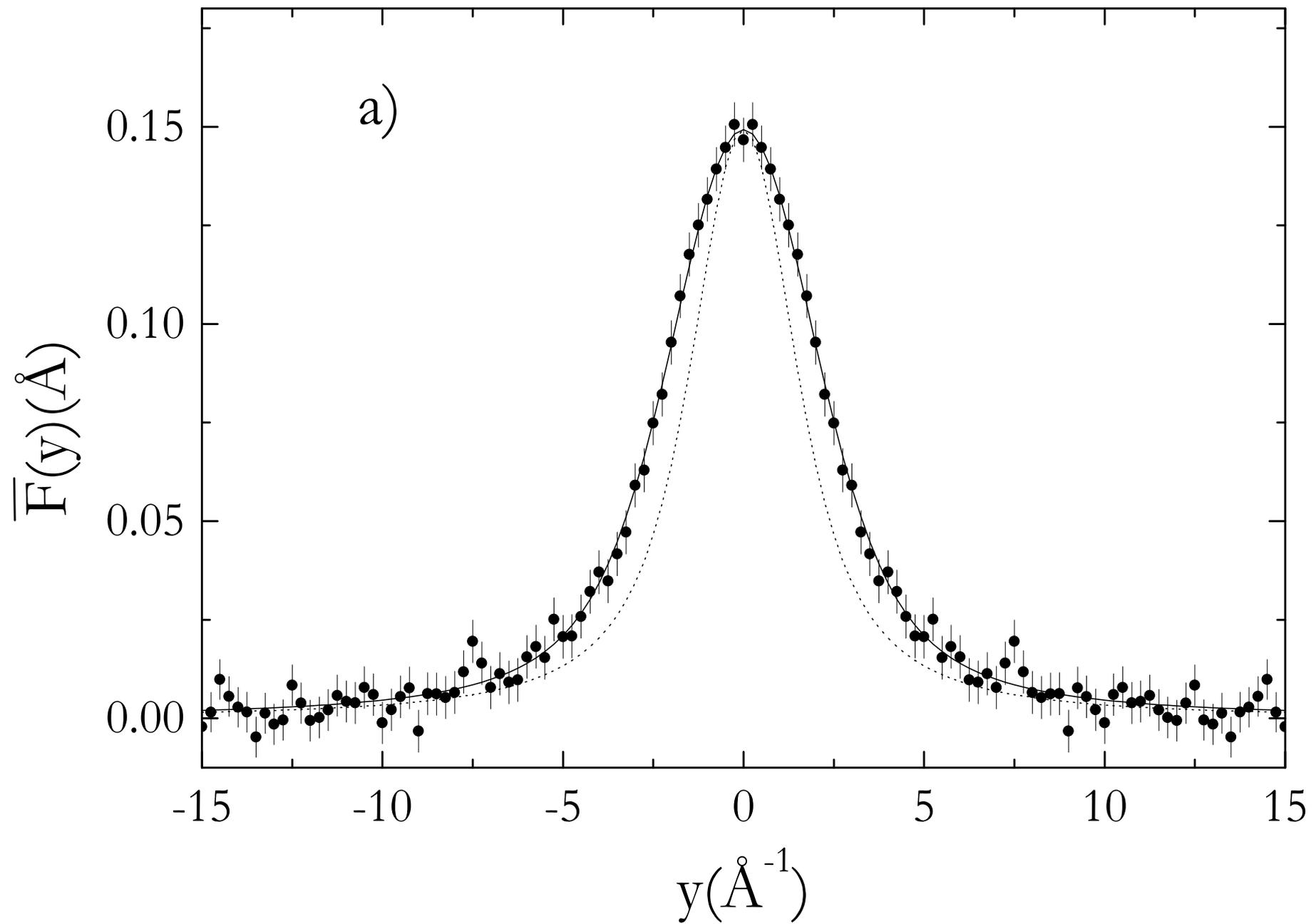
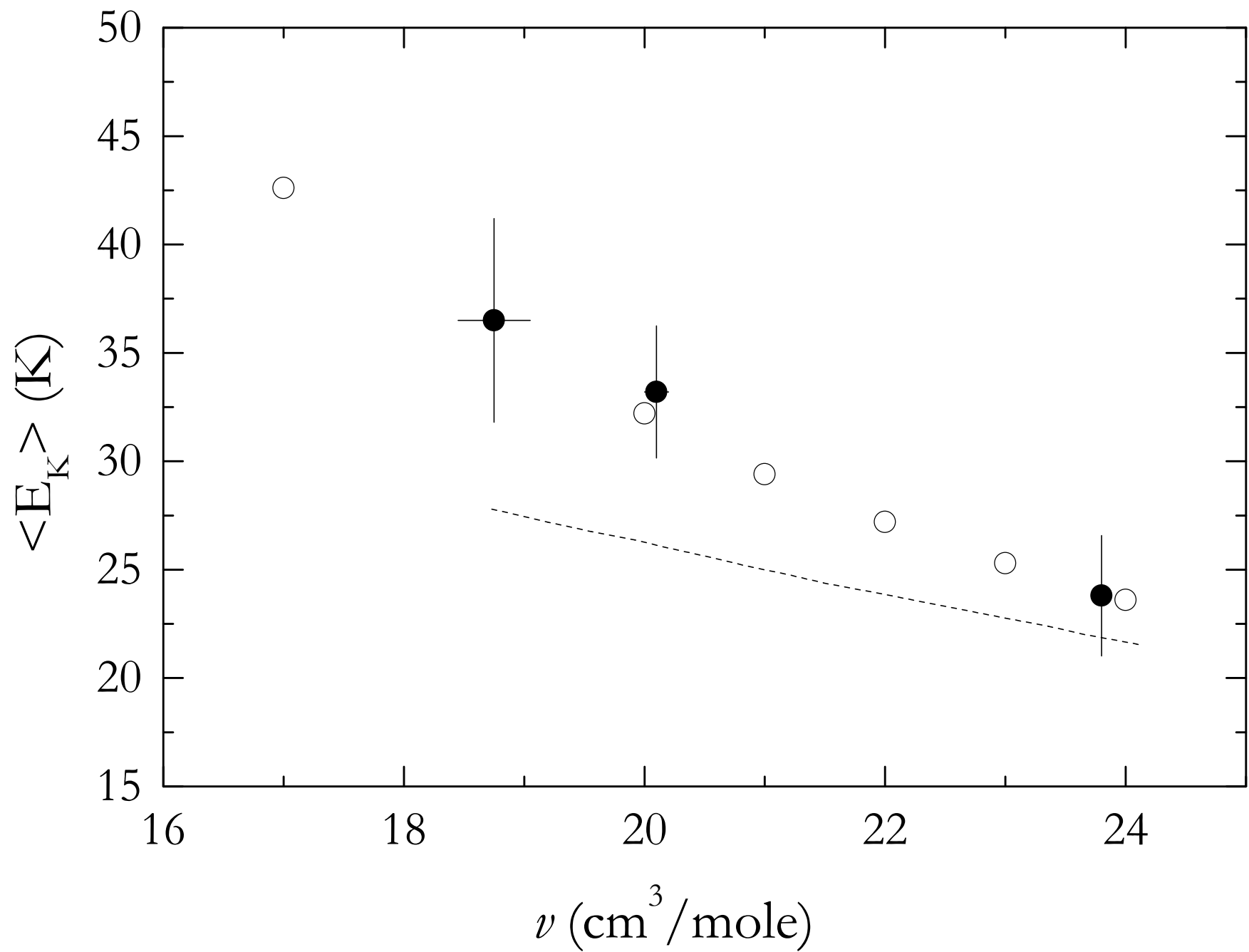
\end{document}